\documentclass[aps,prl,showpacs,superscriptaddress]{revtex4-1}  
\usepackage{hyperref}
\usepackage{graphicx}  
\usepackage{amsmath} 
\usepackage{amssymb} 
\usepackage{color}

\usepackage{dcolumn}
\usepackage{bm}
\usepackage{color}

\usepackage{tabularx}
\usepackage{array}
\newcolumntype{L}[1]{>{\raggedright\let\newline\\\arraybackslash\hspace{0pt}}m{#1}}
\newcolumntype{C}[1]{>{\centering\let\newline\\\arraybackslash\hspace{0pt}}m{#1}}
\newcolumntype{R}[1]{>{\raggedleft\let\newline\\\arraybackslash\hspace{0pt}}m{#1}}

\begin{document}

\title{Assessing diversity in multiplex networks}

\author{Laura C. Carpi}
\affiliation{Programa de P\'os-Gradua\c c\~ao em Modelagem Matem\'atica e Computacional, PPGMMC, Centro Federal de Educa\c c\~ao Tecnol\'ogica de Minas Gerais, CEFET-MG. Av. Amazonas, 7675. 30510-000. Belo Horizonte, MG, Brazil}%

\author{Tiago A. Schieber}
\affiliation{%
Departmento de Engenharia de Produ\c c\~ao, \\Universidade Federal de Minas Gerais, Belo Horizonte, MG, Brazil}%

\author{Panos M. Pardalos}
\affiliation{Industrial and Systems Engineering, University of Florida, Gainesville, FL, USA }%

\author{Gemma Marfany}
\affiliation{Departament de Gen\`etica, Microbiologia i Estadística, Facultat de Biologia, Universitat de Barcelona, Barcelona, Spain}
\affiliation{Institut de Biomedicina de la Universitat de Barcelona (IBUB-IRSJD), Barcelona, Spain}

\author{Cristina Masoller}
\affiliation{Departament de F\'isica, Universitat Polit\`ecnica de Catalunya. Rambla St. Nebridi 22, Terrassa 08222, Barcelona, Spain}%

\author{Albert D\'iaz-Guilera}
\affiliation{
Departament de F\'isica Fonamental, Universitat de Barcelona, Barcelona, Spain
}%
\affiliation{Universitat de Barcelona, Institute of Complex Systems (UBICS), 08028 Barcelona, Spain}

\author{Mart\'in G. Ravetti}
\email{martin.ravetti@dep.ufmg.br}
 \affiliation{%
Departmento de Engenharia de Produ\c c\~ao, \\Universidade Federal de Minas Gerais, Belo Horizonte, MG, Brazil
}%
\date{\today}

\begin{abstract}

Diversity, understood as the variety of different elements or configurations that an extensive system has, is a crucial property that allows maintaining the system's functionality in a changing environment, where failures, random events or malicious attacks are often unavoidable.
Despite the relevance of preserving diversity in the context of ecology, biology, transport, finances, etc., the elements or configurations that more contribute to the diversity are often unknown, and thus, they can not be protected against failures or environmental crises. This is due to the fact that there is no generic framework that allows identifying which elements or configurations have crucial roles in preserving the diversity of the system. Existing methods treat the level of heterogeneity of a system as a measure of its diversity, being unsuitable when systems are composed of a large number of elements with different attributes and types of interactions. Besides, with limited resources, one needs to find the best preservation policy, i.e., one needs to solve an optimization problem. 
Here we aim to bridge this gap by developing a metric between labeled graphs to compute the diversity of the system, which allows identifying the most relevant components, based on their contribution to a global diversity value. The proposed framework is suitable for large multiplex structures, which are constituted by a set of elements represented as nodes, which have different types of interactions, represented as layers. The proposed method allows us to find, in a genetic network (HIV-1), the elements with the highest diversity values, while in a European airline network, we systematically identify the companies that maximize (and those that less compromise) the variety of options for routes connecting different airports.

\end{abstract}

\maketitle

\section{Introduction}
Diversity is a concept that is frequently used in both, scientific and non-scientific contexts, with the main idea of representing the variety of all different forms a system has.  Diversity refers to populations of elements and applies to rather large systems. Diversity can be classified according to three main characteristics of a population~\cite{Page2011}: diversity in some \textit{attributes} (e.g. atoms with different masses), diversity of \textit{types} (e.g., atoms or molecules), and diversity in \textit{configuration} (e.g., connections between atoms in a molecule). 

Several measures have been proposed in the literature to capture these different aspects of diversity~\cite{Weitzman1992,Bossert2001,Stirling07,barabasi2016,Dehmer2012,Fu2015,Min2014,Mougi2012,Raducha2018,Wang2016}. However, to the best of our knowledge, there is no measure for quantifying, in the context of complex networks, the diversity of the connectivity paths.

In this work, we focus in multiplex networks, which are interconnected layered structures, where each layer is formed by a set of elements, nodes, whose interactions are represented by links~\cite{Kurant2006, Gao2012, Boccaletti2014, Kivela2014}. In these structures the interactions exist only within single layers: node $i$ in layer $p$ is not linked with node $j$ in layer $q$. 

Complex systems with multiplex structures include social systems and transportation systems.  For example, Facebook and Twitter are composed of the same individuals, and an individual in Facebook does not have a direct connection to another individual on Twitter.  As another example, air transportation networks are constituted by airports (the nodes), which are connected by routes (the links) of different airlines (the layers)~\cite{Bargigli2015, Domenico2016, Cantini2015, Gallotti2016, Quattrociocchi2014, Donges2011}. Due to their huge socio-economical impact, a lot of efforts are being focused on understanding the structure and functionality of multiplex networks, by developing appropriated analysis tools \cite{Mucha2010,Bennett2015,Battiston2016,Bianconi2013, Domenico2013,Iacovacci2015,Kleineberg2016,Iacovacci2016}, and characterizing new phenomena emerging due to the layered structure \cite{Cardillo2013,Bashan2013,Radicchi2013,Domenico2016-b,Colak2016,Baggio2016,Kouvaris2016,Requejo2016,Genio2016}. 

In the context of multiplex networks, diversity refers to the variety of connectivity configurations the elements that constitute the network (i.e., the nodes and the layers) have. Why is important to measure the diversity of a multiplex system? Straightforward answers are, to identify and to avoid redundant information (preserving only the information necessary to characterize the system under study properly), and to guarantee the stability of the system, protecting the elements that are critical for maintaining the system functionality, or attacking the elements when the goal is dismantling the network.  
We can consider, for example, the metro transport system of a city, represented by a multiplex network with the metro stations being the nodes, and the different lines that connect them being the layers. The evaluation of the contribution of the different elements to the diversity of this system can identify which stations and/or line sections are crucial for the proper operation of the metro system. Losing diversity will likely cause an overuse of some metro stations and/or lines, generating long delays and even the collapse of the whole system.

Here we propose two measures to quantify diversity at two levels: \textit{local diversity}, which refers to the diversity of the connectivity configurations that a node has in the different layers; and \textit{global diversity}, which refers to how different, regarding connectivity configurations, the layers are. These definitions are inspired by the works of M. Weitzman in 1992 \cite{Weitzman1992} and W. Bossert et al. in 2001 \cite{Bossert2001}. The main idea is that the diversity of a system is defined by the distances between its elements: the larger the distances, the more different the elements are, and the more diverse the system is. Thus, to quantify diversity, it is first necessary to define an appropriate distance between pairs of elements (i.e., nodes or layers). A useful diversity definition requires an appropriate measure of the differences in the connectivity paths.  

We begin by proposing two measures for computing distances between nodes (${\mathcal D_i} (\overline{p},\overline{q})$, referred to as \textit{node difference}, ND) and between layers (${\mathcal D} (\overline{p},\overline{q})$, referred to as \textit{layer difference}, LD). ND quantifies the differences of the connectivity paths of node $i$ in layers $\overline{p}$ and $\overline{q}$, while LD quantifies how different the connectivity paths in layers $\overline{p}$ and $\overline{q}$ are.

Then, we define a \textit{local diversity measure}, $U_i$, that quantifies the diversity of the connectivity paths of a single node in all the layers, and a \textit{global diversity measure}, $U$, that quantifies the diversity of the whole multiplex network. To demonstrate the suitability of these measures we analyze three real-world systems: a social network, a genetic network, and an air transportation network. We also show that quantifying the loss of diversity when removing layers provides a simple solution to the challenging problem of structural reducibility. Additional examples are presented in the Supplementary Information (SI).

\bigskip

\section{Methodology}

\noindent {\bf{Distance measure.}} A multiplex network with $M$ layers, each one with the same set of $N$ nodes, is represented by a set of $M$ ($N\times N$) adjacency matrices, $\mathcal{A}=\{A^{[1]},A^{[2]},\dots,A^{[M]}\}$. From these adjacency matrices, the following probability distributions describing local and global connectivity properties of node $i$ in the different layers, are defined:

1) $\mathcal{N}^{\overline{p}}_i$ is the Node Distance Distribution (NDD) of node $i$ in layer $\overline{p}$: $\mathcal{N}^{\overline{p}}_i(d)$ is the fraction of nodes that are at distance $d$ (shortest path) from node $i$ in layer $\overline{p}$. It provides full information of how node $i$ is connected to all other nodes in layer $\overline{p}$. The set of $N$ NDDs, $\{\mathcal{N}^{\overline{p}}_1, \dots, \mathcal{N}^{\overline{p}}_N\}$, contains  information about the global topology of layer $\overline{p}$, in a compact way, and it was used in~\cite{Schieber2017} to define a distance between unlabeled graphs.
 
2) $\mathcal{T}^{\overline{p}}$ is the Transition Matrix of layer $\overline{p}$: $\mathcal{T}^{\overline{p}}_i(j)$ is the probability that node $j$ in layer $\overline{p}$ is reached, in one step, by a random walker located at node $i$ in $\overline{p}$. $\mathcal{T}^{\overline{p}}$ is the adjacency matrix of layer $\overline{p}$, rescaled by the degree of each node, and contains \emph{local} information about the connectivity in layer $\overline{p}$.

\begin{figure*}[h!]
\includegraphics[width=0.85\linewidth]{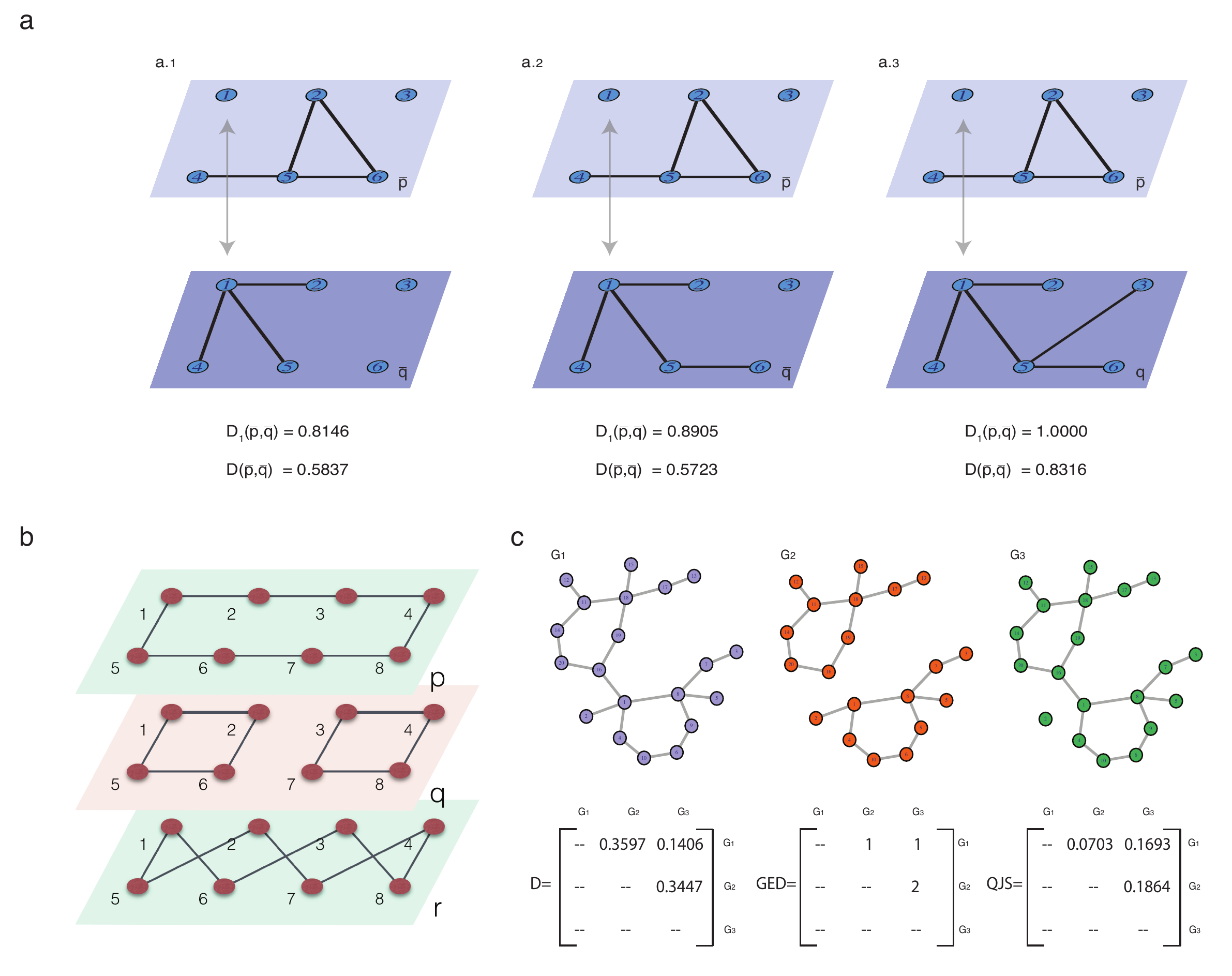}
 \caption{{\bf Example of node difference (ND) and layer difference (LD) calculation.} (a) 
 Dissimilarity of node 1, ${\mathcal D_1(\overline{p},\overline{q})}$, in a three bilayer configurations. ${\mathcal D_1}(\overline{p},\overline{q})$ in panel (a.1) is lower than ${\mathcal D_1}(\overline{p},\overline{q})$ in panel (b.2) because in layer $\overline{p}$ node 1 is disconnected while in layer $\overline{q}$ in (a.1) node 1 has three neighbors, and in (a.2), it has four neighbors. In panel (a.3), ${\mathcal D_1}(\overline{p},\overline{q})$ = 1, because node 1 in layer $\overline{p}$ is disconnected while in $\overline{q}$, it is connected to all nodes. Regarding the layers dissimilarity values, ${\mathcal D(\overline{p},\overline{q})}$, in panel (a.1) the  connectivity of layer $\overline{q}$ is lower than the connectivity of layer $\overline{p}$ (in $\overline{q}$ node 6 is disconnected while in $\overline{p}$, it is connected); in panel (a.3) the connectivity of layer $\overline{q}$ is higher than the connectivity of layer $\overline{p}$ (layer $\overline{q}$ is fully connected while $\overline{p}$ has disconnected nodes). Thus, layers $\overline{p}$ and $\overline{q}$ are more similar in panel (a.2).  
 (b) We consider three layers whose LD values are: ${\mathcal D}{(\overline{p},\overline{q})}=0.4466$, ${\mathcal D}{(\overline{p},\overline{r})}=0.3553$ and ${\mathcal D}{(\overline{q},\overline{r})}=0.6547$. As expected, layers $\overline{q}$ and $\overline{r}$ are the more different (one is disconnected and the other is connected), while $\overline{p}$ and $\overline{r}$ are the more similar (both are connected networks). (c) We consider the graph G1 and analyze the impact of removing a link. We define two networks, G2 is equal to G1, without the link between nodes 1 and 16, and G3 is equal to G1, without the link between nodes 1 and 2. The LD values are ${\mathcal D}{(\overline{G1},\overline{G2})}=0.36$, ${\mathcal D}{(\overline{G1},\overline{G3})}=0.14$ and ${\mathcal D}{(\overline{G2},\overline{G3})}=0.34$. As expected, G1 and G3 are the most similar (they are both connected, and differ only in one link that is peripheral), while G2 is the most different (it is disconnected, the difference with G1 and G3 is one link that is centrally located). Moreover, ${\mathcal D}$ returns similar values for the distance between G1 and G2 and between G2 and G3. In contrast, when using other popular network distances, such as, the Graph Edit Distance \cite{Sanfeliu1983}, and the Quantum Jensen-Shannon divergence \cite{Lamberti2008}, these small or large differences are not well quantified.}

\label{fig:example}
\end{figure*}

Using these distributions, the ND measure of node $i$ in layers $\overline{p}$ and $\overline{q}$ is defined as:
\begin{equation}\label{DI.eq}
{\mathcal D_i} (\overline{p},\overline{q})=\frac{\sqrt{{\cal J} (\mathcal{N}^{\overline{p}}_i,\mathcal{N}^{\overline{q}}_i)} + \sqrt{{\cal J}  (T^{\overline{p}}_{i},T^{\overline{q}}_{i})}}{2 \sqrt{\log{(2)}}},
\end{equation}
where ${\cal J}$ is the Jensen-Shannon (JS) divergence \cite{Cichocki2010} that measures the distance between two probability distributions. With this definition, 
${\mathcal D_i}(\overline{p},\overline{q})=0$ indicates that node $i$ has identical connectivity paths in layers $\overline{p}$ and $\overline{q}$, while ${\mathcal D}_i(\overline{p},\overline{q})=1$ indicates that node $i$ is not connected (or not active) in one layer, while there are paths connecting $i$ to all nodes in the other layer. Figure~\ref{fig:example}-a shows the dissimilarity values of a node, labeled $1$, for three pairs of layers. Node $1$ is not connected in layer $\overline{p}$, while in layer $\overline{q}$, in (a.1), node $1$ is connected with three nodes that posses no further connections; in (a.2) node $1$ is also connected to three nodes, but one of them is also connected with another node; in (a.3), node $1$ is connected, one way or another, to all nodes. These configurations result that ${\mathcal D_1 (a.1)} < {\mathcal D_1 (a.2)} < {\mathcal D_1 (a.3)} = 1$. 
 
The LD measure is the average of ${\mathcal D}_i(\overline{p},\overline{q})$ over all the nodes,
\begin{equation}
{\mathcal D}{(\overline{p},\overline{q})}=\langle{\mathcal D}_i(\overline{p},\overline{q})\rangle _i.
\label{eq:layer_D}
\end{equation}
${\mathcal D}{(\overline{p},\overline{q})}=0$ indicates that layers $\bar{p}$ and $\bar{q}$ are identical, while ${\mathcal D}{(\overline{p},\overline{q})}=1$ indicates that one of the layers is fully connected, while the other is totally disconnected. In Note S1 is presented a discussion regarding the metric properties of ${\mathcal D}$.
 
Figure~\ref{fig:example}-b shows a small example with three labelled k-regular graphs, (all nodes with the same degree), distance ${\mathcal D}$ is able to detect the different connection patterns and the disconnection present in layer $\bar{q}$. In panel c of Figure~\ref{fig:example}, we compare our metric with two others, the Graph Edit Distance (GED) \cite{Sanfeliu1983}, and the Quantum Jensen-Shannon divergence (QSJ) \cite{Lamberti2008}. ${\mathcal D}$ detects and recognizes the important disconnection present in graph G2. Comparisons with other measures are presented in Note S2.

\bigskip

\noindent {\bf Diversity measure.} 
Let $\tilde{S}$ be the set of all entities, which, in the context of multiplex networks, is the set of all nodes, or the set of all layers. Assuming that we have a set $S \subset \tilde{S}$, and that it is possible to compute the distance between all its elements, we define the distance between the element $\overline{g} \not\in S$ and the set $S$, ${\mathcal D}(\overline{g},S)$, as the smallest distance between $\overline{g}$ and any one of the elements of set $S$,  
\begin{equation}
{\mathcal D}(\overline{g},S)= \min_{\overline{s_i} \in S} {\mathcal D}(\overline{g},\overline{s_i}). 
\label{dist_U}
\end{equation}
In other words, the distance of an element that does not belong to the set is defined as the minimum distance of this element, to any element in the set. In Eq.~(\ref{dist_U}), when the population of entities are the nodes,  the distance ${\mathcal D}(\overline{g},\overline{s_i})$ is the node difference (ND), Eq.~(\ref {DI.eq}); while when the entities are the layers, ${\mathcal D}(\overline{g},\overline{s_i})$ is the layer difference (LD), Eq.~(\ref{eq:layer_D}).

Then, the diversity function $U: \tilde{S}\rightarrow \mathbb{R}_+$, is defined recursively as $U(S)=max_{\overline{s_i} \in S} \{U(S\setminus{\overline{s_i}}) + \mathcal D(\overline{s_i},S\setminus{\overline{s_i}})\}$ for all $S \in \tilde{S}$ with $|S| \geq 2$, where $|S|$ represents the cardinality of the set, and $U(S)=0$, for all  $S \in \tilde{S}$  such that $|S|=1$. 

We will use $U_i$ to refer to the diversity of node $i$ in the different layers, and $U$ to refer to the diversity of the set of layers. When an element (a node or a layer) is removed, the diversity of the system decreases. If $\overline{g}$ is removed from $S\cup \overline{g}$, the ``loss'' of diversity is at least equal to $D(\overline{g},S)$, $U(S \cup \overline{g}) \geq U(S) + D(\overline{g},S)$.   The recursion can be optimally solved by using dynamic programming. 

In the case of having all one element sets the same diversity value (zero in this work), it is possible to obtain a diversity ordering, as proposed in Bossert et al. in 2001~\cite{Bossert2001}, where it is also proved the equivalence of these results with those obtained with the Weitzman method~\cite{Weitzman1992}. 

Through a lexicographic-distance based method, a diversity ordering set is obtained, $\mathcal{O}(S)=\{\overline{s_1},\overline{s_2},\dots,\overline{s_{|S|}}\}$, that indicates  the elements in the order of their contribution to the  diversity of the set. 

\begin{figure*}[h!]
\includegraphics[width=\linewidth]{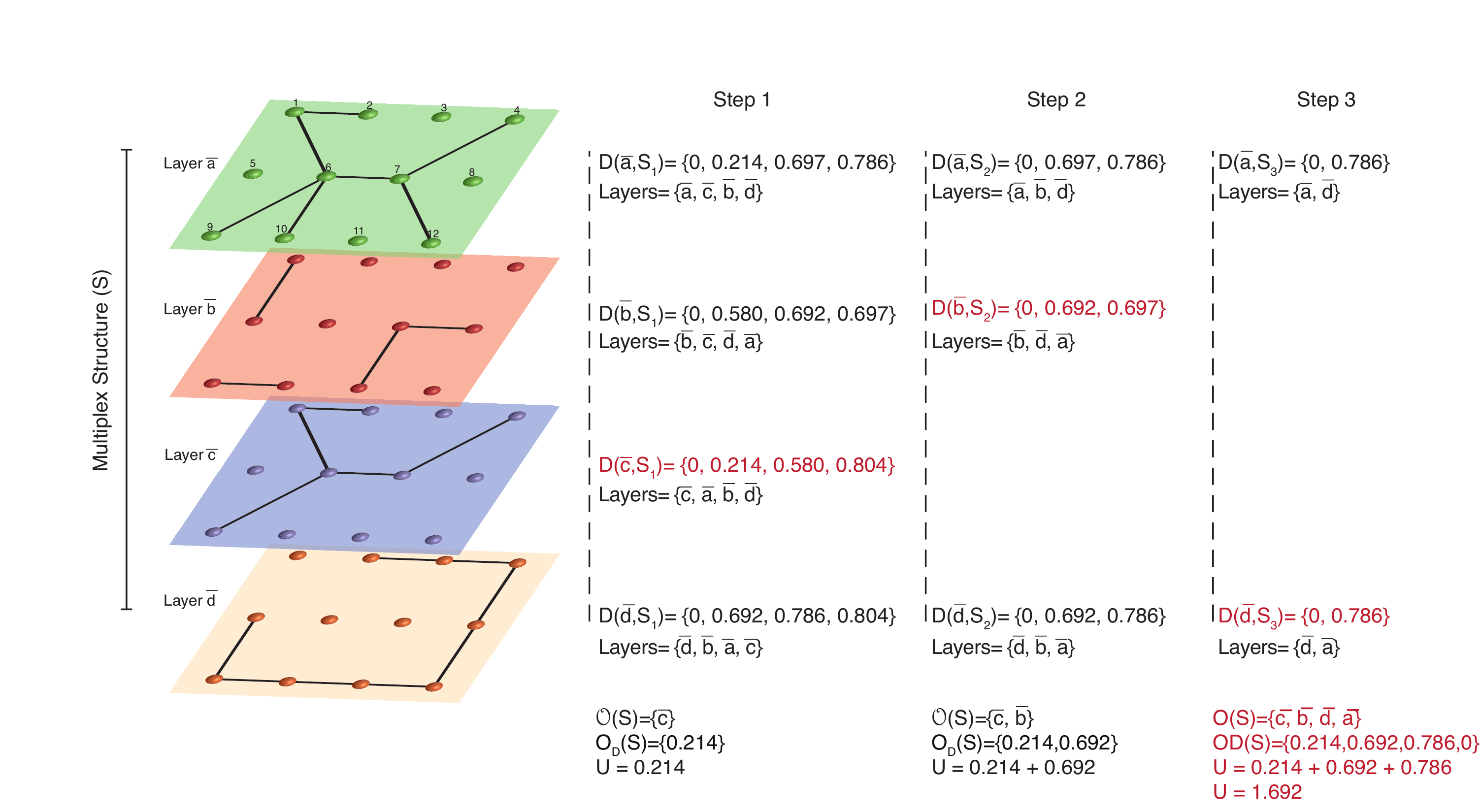}
 \caption{\label{fig:diversityExample} {\bf Computation of the diversity of a 4-layer structure.} The layers have the following LD values: ${\mathcal D}(\bar{a},\bar{b})= 0.697$, ${\mathcal D}(\bar{a},\bar{c})= 0.214$, ${\mathcal D}(\bar{a},\bar{d})= 0.786$, ${\mathcal D}(\bar{b},\bar{c})= 0.580$, ${\mathcal D}(\bar{b},\bar{d})= 0.692$ and ${\mathcal D}(\bar{c},\bar{d})= 0.804$. We calculate the diversity of the system by applying the $U(S)$ equation recursively, with the layers $\bar{s}_i$ ordered accordingly to the distance to the set $S-\bar{s}_i$. Step 1: In $S=\{\bar{a},\bar{b},\bar{c},\bar{d}\}$, layers $\bar{a}$ and $\bar{c}$ present the smallest LD value, and $\bar{c}$ is the layer that less contributes to the diversity of S, as it is closer to the remaining layers . Then, the first step of the recursion gives $U(S)=U(S_1) + 0.214$ where $S_1=S\setminus \bar{c}$. Step 2: In $S_1=\{\bar{a},\bar{b},\bar{d}\}$, layers $\bar{b}$ and $\bar{d}$ present the smallest LD value, and $\bar{b}$ is the layer that less contributes to the diversity of $S_1$. Therefore, $U(S_1)=U(S_2) + 0.692$ where $S_2=S_1\setminus \bar{b}$. Step 3: The LD value of set $S_2=\{\bar{a},\bar{d}\}$ is $\mathcal D(\bar{a},\bar{d})= 0.786$. Since the diversity of a system with cardinality 0 is 0, $U(S_2)={\mathcal D}(\bar{a},\bar{d})$ and $U(S)=0.214+0.692+0.786=1.692$.}
\end{figure*}

We use $U_i(S)$ to refer to the diversity of the connectivity paths that node $i$ has in the different layers, and $U(S)$, to the diversity of the set of layers. Figure~\ref{fig:diversityExample} presents, as an example, the calculation of $U$ for a four-layer network. $U(S)$ is computed through the lexicographical-ordering method, selecting first the less contributing layers to the diversity of the whole system.  

\begin{figure*}[h!]
\includegraphics[scale=0.4]{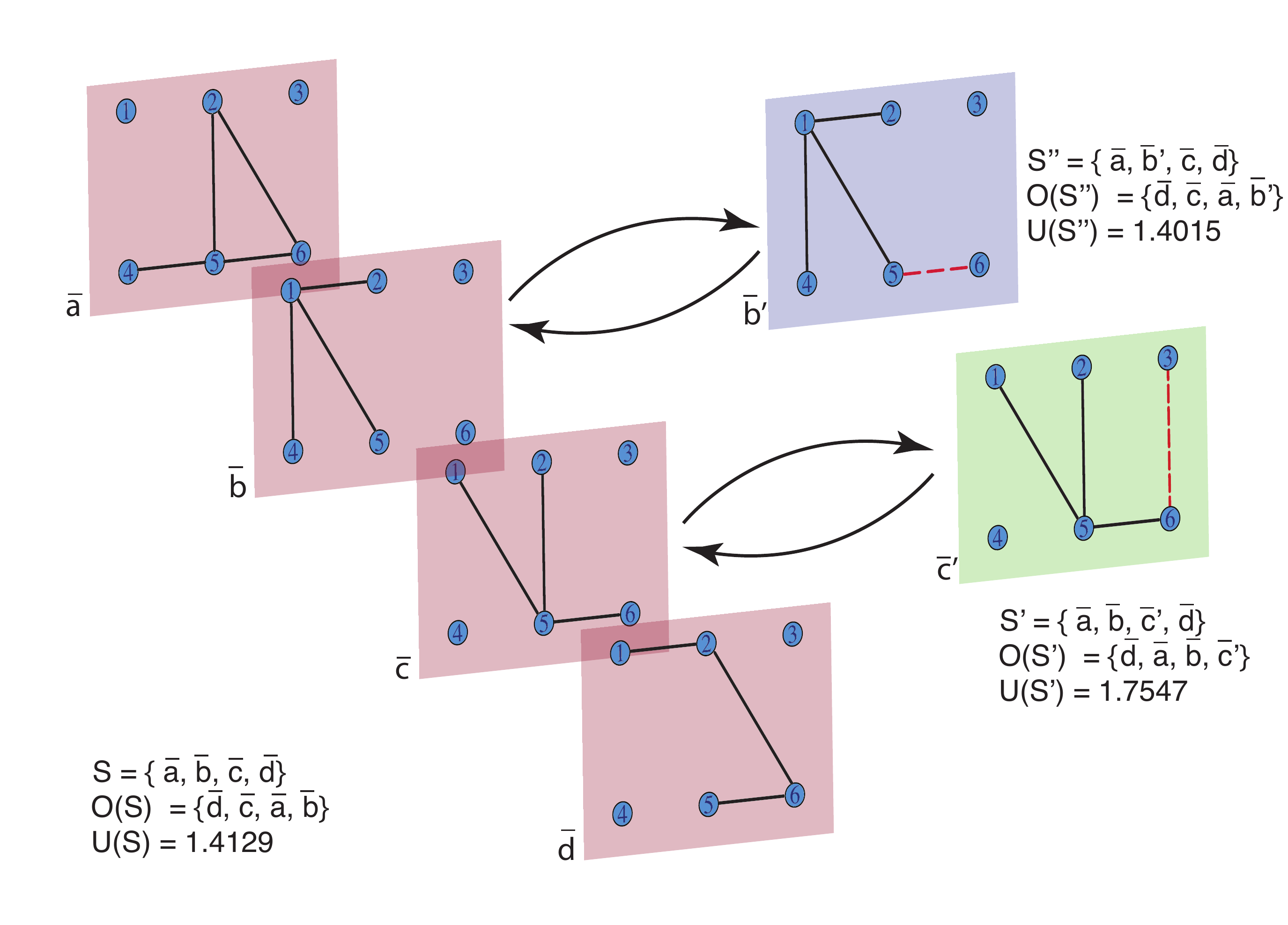}
 \caption{\label{fig:Diversity} {\bf Variation of diversity when adding different links.} Multiplex network $\mathcal{S}$ of 4 layers $\mathcal{S}=\{\bar{a},\bar{b},\bar{c},\bar{d}\}$. In the first exercise, layer $b$ is replaced for layer $b'$ in which link 5-6 is added, and as link 5-6 is present in all the other layers, the diversity of the system decreases. In the second example, layer $c$ is replaced for layer $c'$ in which link 3-6 is added, and, as link 3-6 is not present in the other layers, the diversity of the system increases.}
\end{figure*}

Figure~\ref{fig:Diversity} depicts a multiplex network $\mathcal{S}$ of 4 layers $\mathcal{S}=\{\bar{a},\bar{b},\bar{c},\bar{d}\}$, for which it is computed its ordering set $\mathcal{O}(S)$ and diversity value $U(S)$. Two exercises are performed to show how the diversity measure works. In the first case, layer $b$ is replaced for layer $b'$ in which link 5-6 is added. As link 5-6 is present in all the other layers, and its addition to $b$ does not alter the connectivity of nodes 5 and 6, to the other nodes, the diversity of the system decreases. As the information provided link 5-6 is redundant, the ordering of the elements is not altered.  
In the second example, layer $c$ is replaced for layer $c'$ in which link 3-6 is added. Link 3-6 is not present in the other layers, and its addition creates a connection of node 3 (directly or not), to all other nodes, increasing the diversity of the system. With the addition of link 3-6, the ordering of the elements changes becoming layer $c'$, the one that most contribute to the diversity of the system. All algorithms are freely available at~\footnote{\url{https://github.com/tischieber/assessing_diversity_in_multiplex_networks/}}.


\bigskip
\section{Results}
\noindent {\bf Diversity ordering as a layer reduction method.} 
One direct application of the ordering set is the layer reduction problem, previously studied in~\cite{Manlio2015}. Reducing layers in a multiplex network is a convenient strategy that allows working with a smaller network structure preserving the essential information. In~\cite{Manlio2015} is proposed a method for reducing a multilayer system to another with a smaller number of layers, by using information-theoretic. The method consists in aggregating layers until reaching a balance between the number of layers and the information that must be preserved. 

We propose here the use of the diversity ordering to eliminate layers according to their contribution to the global diversity.
The diversity ordering $\mathcal{O}$ presented above, allows identifying the layer that less contribute to the diversity of the system, guaranteeing the elimination of the most redundant information. The number of layers to be removed depends on the limiting resource, which could be, for example, computational time, money, memory, physical constraints, etc.  In each case, this trade-off has to be analyzed.

The diversity measure quantifies the information lost when reducing the system. As explained before, $U(S \cup \bar{g}) \geq U(S) + {\mathcal D}(\bar{g},S)$, and therefore, when removing a layer, the diversity loss is at least equal to the LD value between the layer removed, and the remaining set.

Figures corresponding to the Human Immunodeficiency Virus (HIV-1) and the European Air Transportation Network (ATN) networks respectively, show how the diversity decreases when removing the layers from the less contributing one. These examples are better analyzed in the corresponding sections. 

\bigskip

\noindent {\bf Measuring diversity in the Aarhus social network.} Here we study the CS-Aarhus Collaboration Network~\cite{Magnani2013} composed by $5$ types of online and offline social interactions between 61 employees including professors, postdoctoral researchers, Ph.D. students and administration staff, at the Department of Computer Science at Aarhus University. The data consists in friendship relationships on Facebook, repeated leisure activities, current working relationships, co-authorship of publications, and regularly eating lunch together between the participants. 

\begin{figure*}[h!]
\includegraphics[scale=0.9]{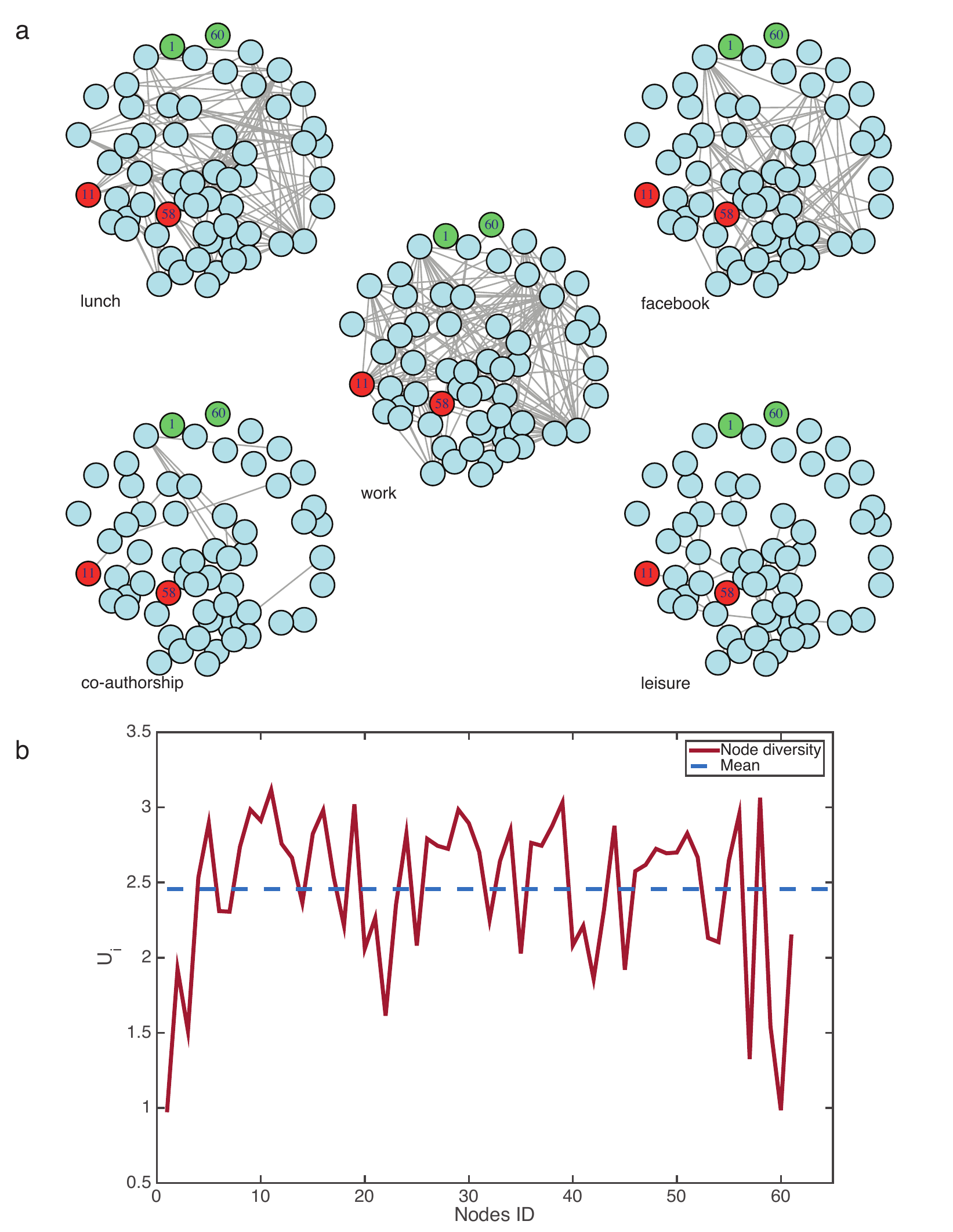}
 \caption{{\bf Analysis of Aarhus social network}. Panel (a) depicts the five layers that constitute the network. The red and green nodes are those whose $U_i$ values are the largest and the lowest, respectively. Panel (b) depicts the diversity, $U_i$, of the nodes. }
\label{fig:Aarhus}
\end{figure*}

Figure~\ref{fig:Aarhus} displays the different layers and the results of the analysis. Most nodes have a similar diversity value $U_i$ (meaning they contribute in a similar way to the diversity of the system), with only two exceptions: nodes ID=1 and ID=60. This is due to the fact that they have very low activity: node $1$ is not active in four layers and has only $2$ connections in the lunch network, while node $60$ is only active in the work network, with $2$ connections.


Considering the LD values, which are listed in Note S3, they reveal that the layers are all very different. However, we can point out some particular results. The co-authorship network, which is the smallest one, is more similar to the Facebook network, meaning that most co-authors are friends on Facebook. The work network is closer to the lunch network, as it can be expected because both are based on routine activities between colleagues working in the same department. The highest LD values correspond to the co-authorship/lunch and co-authorship/work networks, and this is due to the fact that in the co-authorship network most nodes are disconnected, while the lunch and work networks that are very dense. 

\newpage
\noindent {\bf Measuring diversity in the HIV-1 Network.} 

One of the main issues in managing biological networks is the analysis of either huge amount of nodes (i.e. genetic variants in whole genome association analyses) or a limited number of nodes that display complex connections. A useful solution when studying a large set of data is to select fewer representative nodes that recapitulate most of the information of neighboring nodes (for instance, tag SNPs), thus reducing the experimental costs and effort of the analysis. In the second case, casting the data (nodes and links) in the form of a multilayered network provides a useful display of the biological relationships, but the need for pinpointing the relevant informative nodes to simplify the analysis persists. In this context, the measure of node diversity in multiplex networks that we propose becomes a useful criterium to identify biological relevant nodes.  

\begin{figure*}[h!]
\includegraphics[scale=0.45]{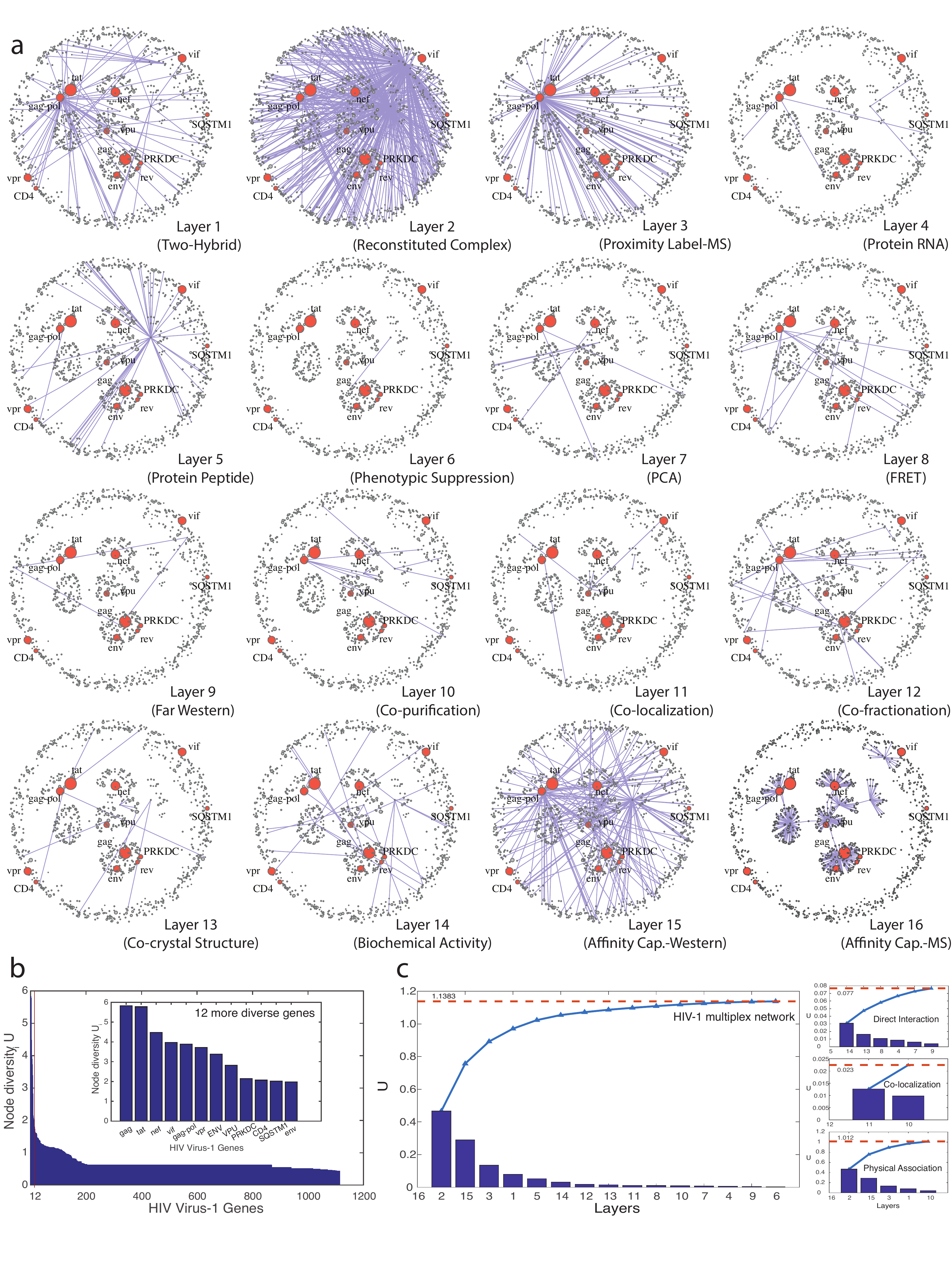}
 \caption{ \label{fig:hiv_final}{\bf Analysis of the multiplex network of the Human HIV-1 virus protein interactions.}  a) Representation of the 16 layers of the multiplex interaction network. The interactions of each layer have been identified using a different experimental technique, as indicated. The 12 most diverse nodes are highlighted (9 out of these 12 nodes are encoded by the HIV-1 genome). Note that these nodes have very heterogeneous connectivity: they are highly connected in some layers, and almost disconnected in others, reflecting different protein properties. b)The plot shows the diversity values of the 1114 nodes. The sub-set of nodes which highest diversity values is indicated. c) Contribution of each layer to the node diversity value Ui.
}

\end{figure*}

To illustrate this concept, we projected the genetic and protein interactions of the human immunodeficiency virus-type 1 (HIV-1) as a multiplex network and analyzed its node diversity. The network consists of 1114 nodes replicated in 16 layers, being the nodes cellular genes and proteins that have been shown to interact with those encoded by HIV-1. Each layers results from the interactions shown by specific types of experimental assays. The data is obtained from BioGRID, a public database that archives and disseminates genetic and protein interactions from humans and model organisms (BioGRID Release 3.4.154, 2017) \cite{biogrid2017} (see Note S4). The diversity values of the 1114 nodes are shown in Figure ~\ref{fig:hiv_final}.  

Our analysis focused first in a reduction of layers and then, in the identification of relevant nodes. Considering that each of the layers results from one experimental assay that explores a particular functional or biochemical trait of each gene/node, and that several assays can be grouped in the same experimental category, our proposed measure of diversity can be used to highlight the most diverse and informative layers within each category in order to reduce the number of relevant layers. In the HIV-cycle network we present here, Physical Association assays include 6 of the layers used in the HIV cycle, namely 1, 2, 3, 10, 15 and 16 (Two hybrid, Reconstituted complex, Proximity label MS, Copurification, Affinity Capture-Western and MS), being 2 and 16 the most diverse layers; Direct Interaction assays include 7 layers, namely 14, 4, 5, 9, 7, 8 and 13 (Biochemical activity, Protein RNA, Protein Peptide, Far Western, PCA, FRET and Co-crystal), being 14 and 5 the most diverse layers; whereas Co-localization assays produce the data of three layers, namely 11, 10 and 12 (Co-localization, co-purification and co-fractionation), being 11 and 12 the most diverse layers. Also, layer 6 (genetic information) is included in other layers and could be omitted because of redundancy. Therefore, by focusing on the more diverse and complementary networks, the multiplex analysis that we propose can reduce the time of analysis of the whole group of functional and interaction assays.

Second, one of the main problems when building biological networks from experimental data is the high variability in the quality and extent of the collected information: high throughput methodologies provide many spurious or irrelevant hits whereas low input approaches are usually too targeted and may miss relevant interactors. As the number of interactors increase, the monolayer networks generated purely by interaction data may become unmanageable. The measure of local diversity that we propose in this work may constitute a very useful curation filter to reduce the network noise produced by both highly promiscuous (high degree) and uninformative (very low degree) nodes, all of which are captured by low diversity Ui values. In this manner, newly aggregated networks built from high diversity nodes selected by different cutoffs may provide relevant insights by focusing on the most informative nodes in specific biological pathways. In this case, most nodes show similar behavior because in almost all the layers they are either active (due to their key role in many cell pathways but not particularly in the HIV cycle) or inactive (reflecting spurious or low relevance hits). These low diversity nodes are captured by low Ui values. On the other hand, a few nodes stand out by their high diversity values, which reflect their heterogeneous participation in the different interaction layers.

Figure ~\ref{fig:hiv_final} (a) displays the 16 layers, and (b) shows the 12 nodes with highest diversity values. It is remarkable that out of the 12 most diverse nodes, 9 are genes encoded by the HIV-1 genome. Particularly, it is worth noticing that two of the most diverse nodes, gag and tat, play complementary roles during the virus infection and harness different cell compartments. Figure ~\ref{fig:hiv_final} (c.1) depicts the layers contribution to the global diversity of the HIV-1 multiplex network. It is possible to see that the pair 2 and 16 brings the 41\% of the global diversity and layer 6 is the less contributing one. Figure ~\ref{fig:hiv_final} (c.2) shows layers contribution to the diversity of the Physical Association assays, (c.3) layers contribution to the Co-localization assay, and (c.4) layers contribution to the Direct Interaction assay.

The node with the highest diversity value, gag, is the main structural precursor protein of HIV-1, which will be processed into four smaller polypeptides with differential functions, from interaction to the lipid cell membrane, interactions with other proteins, RNA binding and activator of the protease and reverse transcriptase functions. Gag has both early and late roles during the HIV-1 life cycle, from capsid assembly and disassembly, envelope protein binding and stabilization, virion maturation after particle release, and early post-entry steps in virus reverse transcription~\cite{Freed1998,Pak2017,Chen2016,Bell2013,Romani2010,Li2016}. In a recent work~\cite{Pak2017}, the gag protein also played an important role in how HIV hacks cells to propagate itself. We highlight that, in the multiplex network structure, gag shows a high diversity value mainly because the experimental approaches to unravel interactions (each layer) target different biochemical traits and thus different type of interactions, in accordance to the diversity of functional domains this precursor protein deploys. 

The second highest diversity node of HIV-1 is tat, a regulatory protein that is key in the 
virus hacking steps to control the cell~\cite{Romani2010,Li2016}. As shown by the interaction nodes in the network, tat is mainly a nuclear protein that interacts with many chromatin remodeling proteins, silencers and activators, which unravel how the virus stealthily takes control of the cell genetic instructions. Besides, tat can also be found in other intracellular compartments and can be secreted extracellularly as a priming factor for HIV-1 infection to other host cells. In fact, tat is the most versatile HIV-1 protein encoded by HIV-1, with different roles depending on its concentration and cell host~\cite{Romani2010}. Therefore, the high diversity value reflects the high diversity of roles, with different interactors unveiled by complementary experimental approaches. Some attempts to gather knowledge on HIV-1 networks relied on the pairwise interactions between the 16 viral proteins as reported in the literature of the last 30 years, and as a result, the gag processed proteins and tat stand out as highly interconnected proteins of the viral genome~\cite{Li2016}. In the multilayer network analysis of the HIV-1 interactions that we propose, gag and tat are also pinpointed as the nodes with the highest diversity, and thus our analysis accurately assesses their pivotal role in the virus and host cell interaction because of their multiplicity of roles. Therefore, we propose that the identification of high diversity nodes in multiplex networks is also useful in the analysis of biological networks.

\bigskip
\noindent {\bf Measuring global diversity in a European Air Transportation Network (ATN).} The network consists of $37$ layers that represent different European airlines, in which nodes are European airports~\cite{Cardillo2013,Estrada2016}. 
We first compute the LD values between all possible pairs of layers, ${\mathcal D}{(\overline{p},\overline{q})}$. 
The lowest ${\mathcal D}$ value corresponds to Iberia/Vueling networks, and indicates that both airlines posses high number of similar routes. 
This is consistent with the fact that Iberia is one of Vueling's main stakeholders, and have a large number of co-shared flights~\footnote{Vueling parent company is IAG, International Airlines Group, an Spanish-English holding created after the fusion of Iberia and British Airways}. Vueling has a hub in Leonardo da Vinci--Fiumicino airport, with an important number of domestic routes in Italy, and LD detects this fact because it places Alitalia as the second airline closest to Vueling's network. 
Ryanair has the highest layer difference value to all other companies. This is because of the use of secondary airports often located away from the center of major towns, such as Warsaw Modlin airport in Prague, and Ciampino airport in Rome, among many others. 
Figure \ref{fig:Vueling} shows the two closest (top) and the two more distant (bottom) companies to Vueling.

\begin{figure*}[h!]
 \includegraphics[width=0.9\textwidth]{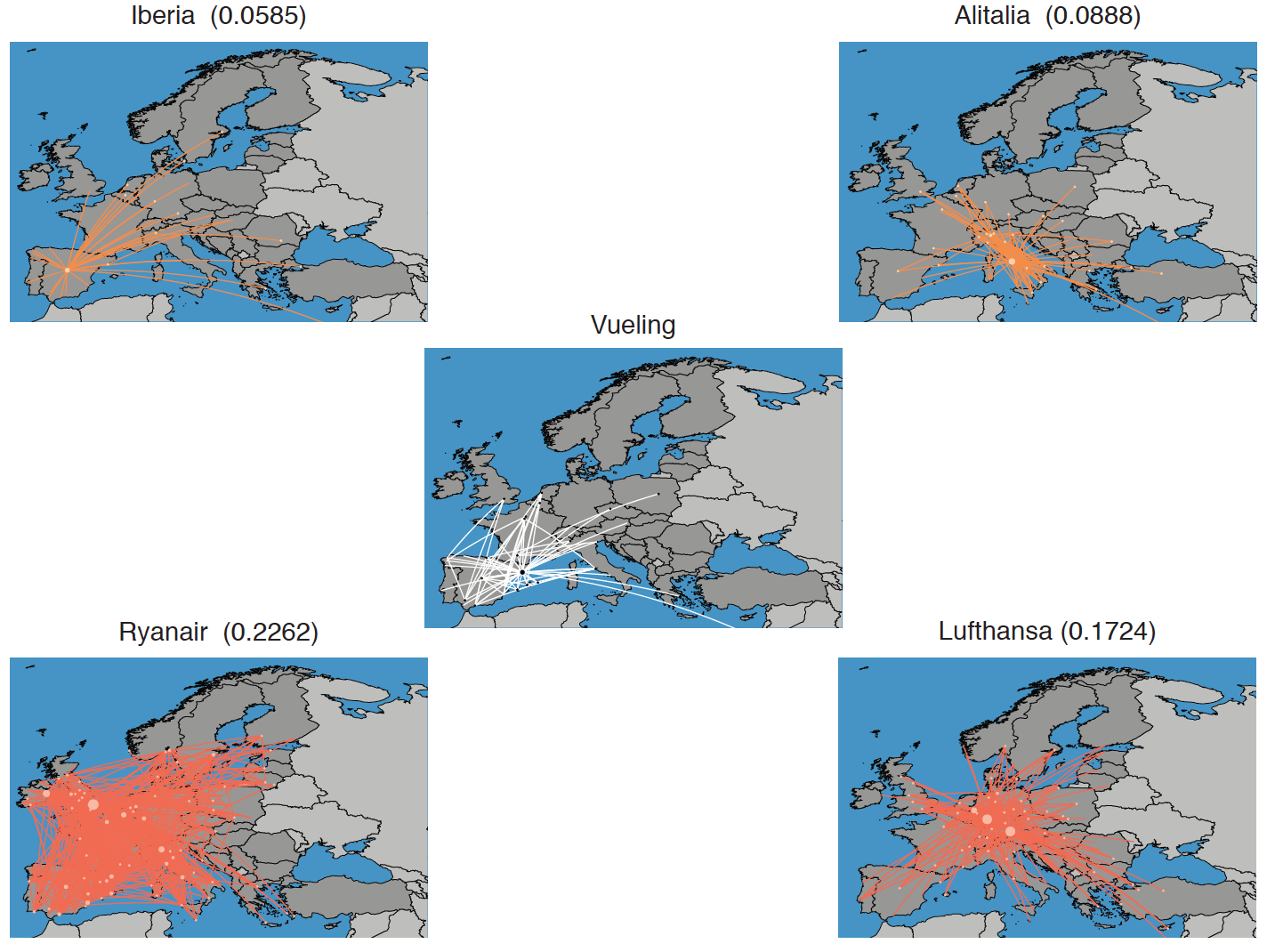}
 \caption{\label{fig:Vueling} {\bf Difference values between Vueling and 4 other companies.} Difference values between Vueling (center), and the two closest (top) and more distant (bottom) networks, Iberia and Alitalia, and Ryanair and Lufthansa, respectively.}
\label{fig:Vueling}
\end{figure*}

We also analyze the diversity of three main air alliances: Star Alliance (SA), One World (OW) and Skyteam (ST). 
Each air alliance is represented as a multiplex network that includes only the European partners, which give $4$ layers for the ST network, $4$ layers for OW, and $9$ layers for SA (see Note S5). 
The efficiency of each alliance relies in covering Europe with the minimum route overlapping. The efficiency defined in these terms is reflected by higher diversity values between the participant airlines. 



\begin{figure*}[ht]
\includegraphics[scale=0.7]{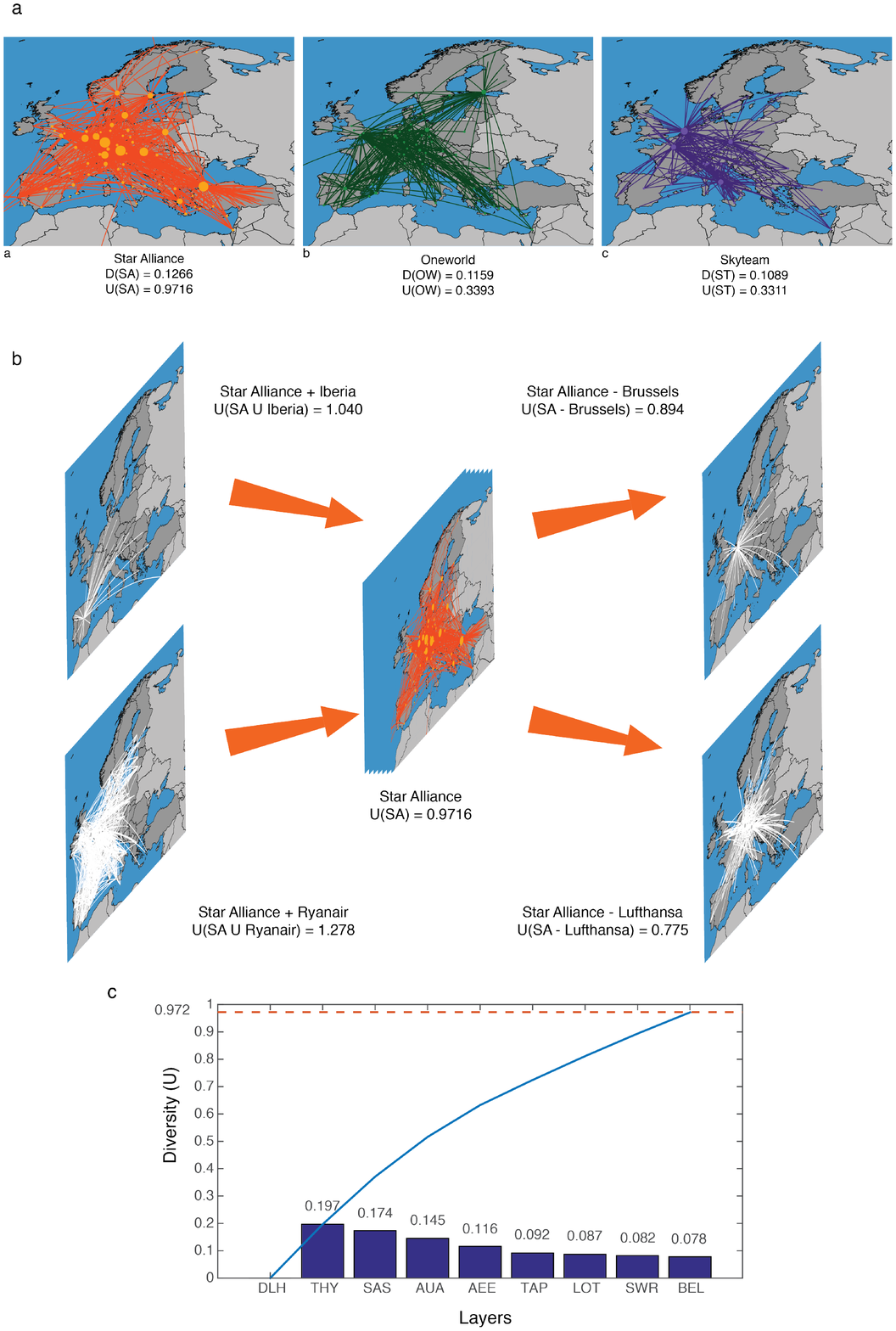}
\caption{\label{fig:Airlines}{\bf Diversity analysis for Alliance groups.} Subfigure (a) shows the aggregated networks of three airlines groups: Star Alliance (SA), One World (OW), and Skyteam (ST), with their correspondent global layer dissimilarity and diversity values. Is important remember that only European companies are considered in this work. Panel (b) presents a diversity analysis of the impact of the individual removal of Lufthansa and Brussels airlines,  and the individual inclusion of Ryanair and Iberia, in Star Alliance. Panel (c) shows the diversity contribution of each airline in the alliance. The blue line shows the global diversity value of the system at each state. The orange dash line indicates the global diversity value of the whole system. }
\end{figure*}

Figure~\ref{fig:Airlines}-a shows the networks of the alliances, with their corresponding LD values and diversity value $U$.  A high $U$ value indicates that the set of routes offered by an alliance are very diverse, implying in a less route overlapping, and wider coverage. Star Alliance is the structure with the highest $U$.  As can be seen in Fig.~\ref{fig:Airlines}-a its coverage and spanning through European airports is visibly higher, at least for the European airlines considered in the analysis.


As second application, we consider only the Star Alliance group. In this data set, SA has 9 European companies, with a diversity value $U(SA)= 0.9716$. The diversity ordering is $\mathcal{O}$(SA)=\{\textit{Brussels Airlines (BEL), Swiss Air (SWR), Polish Airlines (LOT), Air Portugal (TAP), Aegean Airlines (AEE), Austrian Airlines (AUA), Scandinavian Airlines (SAS), Turkish Airlines (THY), Lufthansa (DLH)}\}.

This ordering uncovers some very interesting facts. Lufthansa is the company that brings more diversity to the alliance; its removal causes the greatest drop in the diversity value. 
On the contrary, the removal of Brussels Airlines causes the smallest impact to diversity.  It is interesting noticing that some small companies, like Aegean Airlines, highly contribute to the diversity as they cover routes not attended by their partners. When considering companies from outside the group, we observe that the inclusion of Iberia less increases the diversity of the group, and Ryanair is the one that would contributes the most (see Figure~\ref{fig:Airlines}-b). It is important to note that we did not aggregate the airports into cities, thus, the measure considers that arriving into two different airports in the same city are different destinations. For the diversity exercise, this is not relevant, however, it will be interesting, as future work, to take this into account, as aggregating the airport that serve the same city could have important implications for the economic interpretation of the results.
Figure \ref{fig:Airlines}-c shows the diversity loss when removing layers from those that less contribute, to those that more contribute to the diversity of the Star Alliance (SA) network. Here we see that the maximum diversity value corresponds to 9 layers. As previously described, the ordering set gives the sequence of the layers according to their contribution to the global diversity. In the case of the SA network, if Brussels Airlines is removed (less contributing layer), the diversity decreases by 8\%. As we continue to remove layers following the ordering $\mathcal{O}(SA)$, the diversity gradually decreases until the system is reduced to just one layer (Lufthansa).
Experiments with other real networks can be found in Note S6.

\bigskip
\section{Conclusions}
We have proposed a new approach for quantifying the diversity of a multiplex system. First, we have defined metric distances between nodes and layers, which were then used to define the diversity of the connectivity paths of a node in the different layers, and the diversity of the connectivity paths of the whole set of layers. We have applied these measures to study real-world networks (the Aarhus social network, the HIV-1 network, and the European air-traffic network), and we have also used them for optimizing layer reduction, with minimum diversity loss, while avoiding the creation of layers which do not exist. 

In the HIV-1 genetic network, the nodes with highest diversity values unveiled by our analysis are those that play more versatile roles in the life-cycle of the virus and show higher interplay with the host cell proteins, thus being key to pathogenesis. 
The analysis of the air-traffic network detected which airlines, when joining an alliance, optimally increase the diversity of the structure, bringing new routes while minimizing overlapping ones; and which ones, when leaving the alliance, less compromise the diversity of the routes offered by the group. 

The measures proposed here can have important practical applications for urban transportation systems, ecological systems, financial systems, etc. They can be used to detect which elements need to be modified in order to increase, or to decrease, the diversity of the system. 

\section{Acknowledgments}
Research partially supported by Brazilian agencies FAPEMIG, CAPES, and  CNPq. 
P.M.P. acknowledges support from the ``Paul and Heidi Brown Preeminent Professorship in ISE, University of Florida'', and RSF 14-41-00039. G.M. acknowledges financial support from MINECO, project SAF2016-80937 and from Generalitat de Catalunya Project 2014SGR-0932.
C. M. acknowledges
partial support from Spanish MINECO (FIS2015-
66503-C3-2-P) and ICREA ACADEMIA. 
A.D.-G. acknowledges 
financial support from MINECO, Project FIS2015-71582, and from Generalitat
de Catalunya Project 2014SGR-608. 
M.G.R. acknowledges
partial support from FUNDEP.

\bigskip
\noindent {\bf Additional Information}\\
The authors declare no competing interests.


\noindent {\bf Author contributions statements}\\
L.C., T.S. and M.G.R. planned the experiments, implemented and tested the algorithms. L.C.,T.S, A.D-G, C.M, P.P. and M.G.R. analyzed the results and wrote the article. G.M. analyzed and wrote the HIV experiment.
All authors revised the whole manuscript.

\clearpage

\end{document}